\documentstyle[12pt]{article}
\input amssym.def
\input amssym.tex

\def\goth{\frak}
\def\double{\Bbb}
\def\ccal{\cal}

\def\cc{{\double C}}

\def\rr{{\double R}}

\def\aa{{\cal A}}

\def\dd{{\cal D}}
\def\gg{{\goth g}}
\def\hh{{\cal H}}
\def\hhh{{{\double H}}}

\def\mm{{{\ccal M}}}

\def\aa{{\cal A}}
\def\dd{{\cal D}}
\def\hh{{\cal H}}

\def\t{\,{\rm tr}\,}

\def\ot{\otimes}
\def\op{\oplus}

\def\bb{\begin{eqnarray}}
\def\ee{\end{eqnarray}}
\def\eee{\nonumber\end{eqnarray}}
\def\pp{\pmatrix}
\def\qq{\quad}
\def\del{\partial}

\begin{document}

\hsize 17truecm
\vsize 24truecm
\font\twelve=cmbx10 at 13pt
\font\eightrm=cmr8
\baselineskip 18pt

\begin{titlepage}

\centerline{\twelve CENTRE DE PHYSIQUE THEORIQUE}
\centerline{\twelve CNRS - Luminy, Case 907}
\centerline{\twelve 13288 Marseille Cedex}
\vskip 4truecm

\centerline{\twelve  Fuzzy Mass Relations in the
Standard Model}

\bigskip

\begin{center}
{\bf Bruno IOCHUM}
\footnote{ and Universit\'e de Provence,
\qq\qq iochum@cpt.univ-mrs.fr
\qq schucker@cpt.univ-mrs.fr} \\
\bf Daniel KASTLER
\footnote{ and Universit\'e d'Aix-Marseille II} \\
\bf Thomas SCH\"UCKER $^{1}$
\end{center}

\vskip 2truecm
\leftskip=1cm
\rightskip=1cm
\centerline{\bf Abstract}

\medskip

Recently Connes has proposed a new geometric version
of the standard model including a non-commutative
charge conjugation. We present a systematic analysis
of the relations among masses and coupling constants
in this approach. In particular, for a given top mass,
the Higgs mass is constrained to lie in an interval.
Therefore this constraint is locally stable under
renormalization flow.

\vskip 1truecm
PACS-92: 11.15 Gauge field theories\\
\indent
MSC-91: 81E13 Yang-Mills and other gauge theories

\vskip 2truecm

\noindent july 1995
\vskip 1truecm
\noindent CPT-95/P.3235\\
\noindent hep-th/9507150

\vskip1truecm

 \end{titlepage}

\section{Introduction}

In his beautiful book \cite{cbook}, A. Connes applies
non-commutative geometry to the standard model of
particles. The last theorem of this book states that the
ordinary Lagrangian of the standard model with three
generations of leptons and quarks and one doublet of
Higgs scalars has a natural algebraic interpretation.
Its principal ingredients are two algebras and a
generalized Dirac operator. From these,
Connes constructs two differential algebras, two gauge
potentials, their curvatures and the Euclidean
Yang-Mills actions as  scalar products of the
curvatures with themselves. When applied to the
commutative case --- the commutative algebra of
smooth functions on a four dimensional spacetime and
the genuine Dirac operator --- this  Yang-Mills action
and the covariantized Dirac action reproduce spinor
electrodynamics. However, when applied to the tensor
product of the commutative spacetime algebra with two
non-commutative internal algebras these two
Lagrangians reproduce
exactly the Lagrangian of the
standard model including the entire Higgs sector, i.e.
the Klein-Gordon Lagrangian for the Higgs scalars,
their Higgs potential and their Yukawa terms. In
particular, the 18 free parameters of the standard
model (which can be taken to be the three gauge
couplings, $g_1,\ g_2,\ g_3$, the masses of the $W$,
 of the Higgs, of three leptons and of six quarks, and
four mixing parameters in the
Cabbibo-Kobayashi-Maskawa matrix) remain free and
are the only free physical parameters in the
non-commutative approach.

In its original version, this non-commutative
approach due to A. Connes and J. Lott
\cite{cl},\cite{ort} had two major shortcomings: the
need of two algebras with related bimodules and two
extra $U(1)$ factors in the gauge group that had to be
eliminated by two additional algebraic (unimodularity)
conditions. Recently, Connes \cite{creal} has improved
this framework by introducing a real structure of a
spectral triplet $(\aa,\hh,\dd)$, where $\aa$ is a {\it
real} algebra represented on the Hilbert space $\hh$
and $\dd$ a Dirac operator on $\hh$. The real structure
is given by an anti-unitary operator $J$ on $\hh$
which, in commutative geometry, is the charge
conjugation. Now, the internal space of the standard
model is described by {\it one} algebra
$\aa=\hhh\op\cc\op M_3(\cc)$, $\hhh$ is the algebra
of quaternions, the Hilbert space is spanned by all
leptons and quarks and the Dirac operator is given by
the fermionic mass matrix. The hypothesis of the
quoted theorem now becomes extremely simple. The
gauge group $G$ is the group of unitaries of $\aa$,
$G=SU(2)\times U(1)\times U(3)$.The gauge potential
is a 1-form in {\it the} differential algebra of the triple
$(\aa,\hh,\dd)$ and its Yang-Mills action together
with the {\it Majorana} action suitably
covariantized yields the
action of the standard model with a doublet of Higgs
scalars. There are again the 18 free
parameters.

In the standard model, the space of
parameters is a direct product of 18 intervals. In the
non-commutative approach, the space of parameters
which, according to Connes' theorem, has a non-empty
interior, reveals an interesting shape. In particular,
$\sin^2\theta_w$ is bounded from above, the mass of
the $W$ lies essentially between the lightest and the
heaviest fermion mass and the Higgs mass is bounded
from below and above. Since all values of the interval
are possible, we call this framing a {\it fuzzy mass
relation}.

 Throughout this paper, we  assume that all
 fermion masses are different and
that the Cabbibo-Kobayashi-Maskawa matrix is
non-degenerate, i.e. has no proper invariant subspace.

\begin{description}
\item{\bf Theorem.}
{\it
\begin{enumerate}
\item[i.]
\bb\sin^2\theta_w<\frac{2}{3}\,\left(1+\frac{1}{9}
\left(\frac{g_2}{g_3}\right)^2\right)^{-1}.
\label{sinth}\ee
\item[ii.]
If the heaviest lepton $\tau$ satisfies
$m_\tau^2<(m_t^2+m_b^2+m_c^2+m_s^2+m_u^2+m_d^2)
/3$, then, with $e$ the lightest lepton, we have
\bb m_e^2<m_W^2<
(m_t^2+m_b^2+m_c^2+m_s^2+m_u^2+m_d^2)/3.
\label{wth}\ee
\item[iii.]
\bb m_{H\,min}^2<m_H^2<m_{H\,max}^2 \label{hth}\ee
where $m_{H\,min}^2$ and $m_{H\,max}^2$ depend
on all fermion masses but $m_\mu$. The bounds are
given by equations (\ref{higgsbounds}) and
(\ref{tildeh}) and plotted in the figure.
\hfil\break\noindent
  $m_{H\,max}^2-m_{H\,min}^2$
factorizes $(m_\tau^2-m_e^2)$ and
$(m_t^2+m_b^2+m_c^2+m_s^2+m_u^2+m_d^2-3m_W^2)$.
In particular, neglecting all fermion masses but
$m_\tau$ and $m_t$, we have
\bb m_{H\,max}-m_{H\,min}=\left[k\left(
\frac{m_\tau}{m_t}\right)^2+O\left(\left(
\frac{m_\tau}{m_t}\right)^4\right)\right]\,m_t\eee
where $k$ is given in equations (\ref{kon}-\ref{r})
and is of order one for experimental values of $m_W$
and $m_t$. \end{enumerate}
}
\end{description}

\section{The geometric version of the standard model}

The basis of non-commutative geometry is a (real)
spectral triple $(\aa,\hh,\dd)$.

$\aa$ is a real,
associative algebra with unit 1 and involution $^*$.
The spacetime $M$ is described by the infinite
dimensional commutative algebra of smooth
functions $f:M\longrightarrow \cc$ with involution
$f^*=\bar f$, the complex conjugate. The internal
space is described by a finite dimensional algebra
whose group of unitaries $ G: = \{g\in \aa\ |\ gg^\ast
=g^*g=1\}$ will contain the gauge group. In the case
of the standard model, this choice is
\bb \aa=\hhh\op\cc\op M_3(\cc)\qq{\rm with}\qq
G=SU(2)\times U(1)\times U(3). \eee
 We denote by $\hhh$ the
algebra of quaternions, viewed as $2\times 2$
matrices,
\bb \pp{x&-\bar y\cr y&\bar x}\,\in\hhh,\qq
x,y\in\cc.\eee

$\hh$ is a Hilbert space carrying a faithful
representation $\rho$ of the algebra $\aa$. We also
assume that $\hh$ is equipped with a chirality $\chi$
and a charge conjugation $J$. The chirality is a
unitary operator of square one that commutes with the
representation. Therefore $\chi$ decomposes the
representation space
 into a left-handed piece $(1-\chi)/2\,\hh$ and a
right-handed piece $(1+\chi)/2\,\hh$.
The charge conjugation is an anti-unitary operator of
square plus or minus one, depending on spacetime
 dimension and signature. Also depending on
spacetime dimension and signature, $J$ commutes or
anticommutes with $\chi$.
We further assume that
\bb\bullet\qq \rho(a)\ {\rm commutes\
with}\   J\rho(\tilde a)J^{-1},\ \ {\rm for\ all}\
a,\tilde a\ {\rm in}\ \aa.\qq\qq\qq\qq\qq\qq\qq\qq
\qq\qq\qq\ \label{oo}\ee
The charge conjugation as well decomposes the
representation space into two pieces, particles and
anti-particles,
\bb \hh=\hh_L\op\hh_R\op\hh_L^c\op\hh_R^c.\eee

For a four dimensional
spacetime, the Hilbert space consists of all square
integrable (Dirac) spinors, a function $f$ acting on a
spinor $\psi$ by multiplication,
$(\rho(f)\psi)(x):=f(x)\psi(x)$. The chirality
$\chi=\gamma_5$ decomposes a Dirac spinor into left-
and right-handed (Weyl) spinors and the charge
conjugation acts on $\psi$ as
$\psi^c:=i\gamma^2\bar\psi$ where $\gamma^2$ is
the second Dirac matrix and the bar denotes complex
conjugation. The internal space counts as zero
dimensional \cite{cbook}.
 Its Hilbert space is finite dimensional and
contains all fermions. For the standard model, we have
\bb\hh_L=
\left(\cc^2\ot\cc^N\ot\cc^3\right)\ \op\
\left(\cc^2\ot\cc^N\ot\cc\right),\eee
\bb\hh_R=\left((\cc\op\cc)\ot\cc^N\ot\cc^3\right)\
\op\ \left(\cc\ot\cc^N\ot\cc\right).\eee
 In each summand, the first factor
denotes weak isospin doublets or singlets, the second
$N$ generations, $N=3$, and the third denotes colour
triplets or singlets.
\hfil\break\noindent
Let us choose the following basis
of
$\hh=\cc^{90}$:
\bb
& \pp{u\cr d}_L,\ \pp{c\cr s}_L,\ \pp{t\cr b}_L,\
\pp{\nu_e\cr e}_L,\ \pp{\nu_\mu\cr\mu}_L,\
\pp{\nu_\tau\cr\tau}_L;&\cr \cr
&\matrix{u_R,\cr d_R,}\qq \matrix{c_R,\cr s_R,}\qq
\matrix{t_R,\cr b_R,}\qq  e_R,\qq \mu_R,\qq
\tau_R;&\cr  \cr
& \pp{u\cr d}^c_L,\ \pp{c\cr s}_L^c,\
\pp{t\cr b}_L^c,\
\pp{\nu_e\cr e}_L^c,\ \pp{\nu_\mu\cr\mu}_L^c,\
\pp{\nu_\tau\cr\tau}_L^c;&\cr\cr
&\matrix{u_R^c,\cr d_R^c,}\qq
\matrix{c_R^c,\cr s_R^c,}\qq
\matrix{t_R^c,\cr b_R^c,}\qq  e_R^c,\qq \mu_R^c,\qq
\tau_R^c.&\eee
Let $(a,b,c)\in\hhh\op\cc\op M_3(\cc)$ be an
element in the algebra $\aa$. $\rho$ acts on the above
Hilbert space by
\bb \rho(a,b,c):=\pp{
\rho_w(a,b)&0\cr
0&\bar\rho_s(b,c)}\eee
with
\bb\rho_w(a,b)&:=&\pp{
a\ot 1_N\ot 1_3&0&0&0\cr
0&a\ot 1_N&0&0\cr
0&0&B\ot 1_N\ot 1_3&0\cr
0&0&0&\bar
b1_N},\qq
B:=\pp{b&0\cr 0&\bar b},
\cr  \cr \cr
  \rho_s(b,c)&:=&\pp{
1_2\ot 1_N\ot c&0&0&0\cr
0&\bar b1_2\ot 1_N&0&0\cr
0&0&1_2\ot 1_N\ot c&0\cr
0&0&0&\bar b1_N}.
\eee
The chosen representation $\rho$ will take into
account weak interactions $\rho_w(a,b),\ a\in\hhh,\
b\in\cc$, and strong interactions $\rho_s(b,c),\ c\in
M_3(\cc)$, $c$ for colour. This choice discriminates
between leptons (colour singlets) and quarks (colour
triplets). The role of $b\in\cc$ appearing in both
weak interactions $\rho_w(a,b)$ and strong
interactions $\rho_s(b,c)$ is crucial to make
$\rho(a,b,c)$ a representation of $\aa$ and is crucial
for weak hypercharge computations. There is an
apparent asymmetry between particles and
anti-particles, the former are subject to weak, the
latter to strong interactions. However, since particles
and anti-particles are permuted by $J$ via the
fundamental property (\ref{oo}), the theory is
invariant under charge conjugation.

The chirality operator and charge conjugation are
\bb\chi=\pp{
-1_{24}&0&0&0\cr
0&+1_{21}&0&0\cr
0&0&-1_{24}&0\cr
0&0&0&+1_{21}},\qq J=\pp{0&1_{45}\cr 1_{45}&0}C,\eee
 $C$ being the complex conjugation.

The last item in the spectral triple is the
(generalized) Dirac operator $\dd$, a selfadjoint
operator with the following properties:
\begin{itemize}
\item
$ \dd\chi=-\chi\dd,$
\item
$\dd J=+J\dd,$
\item
$[\dd,\rho(a)]$ is bounded for all $a$ in $\aa$,
\item
$[\dd,\rho(a)]$ commutes with  $J\rho(\tilde a)J^{-1}$,
for all
$a,\tilde a$ in $\aa$.
\end{itemize}
For spacetime, $\dd$ is the genuine Dirac operator.
For the internal space, $\dd$ is made up with the
fermionic mass matrix $\mm$,
\bb\dd=\pp{
0&\mm&0&0\cr
\mm^*&0&0&0\cr
0&0&0&\mm\cr
0&0&\mm^*&0}.\eee
Let us recall the mass matrix of the standard model:
\bb\mm=\pp{
\pp{M_u\ot1_3&0\cr 0&M_d\ot 1_3}&0\cr
0&\pp{0\cr M_e}},\eee
with
\bb M_u:=\pp{
m_u&0&0\cr
0&m_c&0\cr
0&0&m_t},\qq M_d:= C_{KM}\pp{
m_d&0&0\cr
0&m_s&0\cr
0&0&m_b},\qq M_e:=\pp{
m_e&0&0\cr
0&m_\mu&0\cr
0&0&m_\tau}.\eee
 All indicated fermion masses are supposed positive and
different. The
Cabbibo-Kobayashi-Maskawa matrix  $C_{KM}$ is
supposed non-degenerate in the sense that there is no
simultaneous mass and weak interaction
eigenstate.
\hfil\break\noindent
Note that the
 strong interactions are vector-like: for all $b\in\cc$
and $c\in M_3(\cc)$,  $\rho_3(b,c)$ commutes with
the corresponding restrictions of $\chi$ and $\dd$.

A last ingredient of the general theory is another
operator $z$ on the Hilbert space.
$z$ is used to construct a gauge invariant scalar
product $(\omega,\kappa):=\t(\omega^*\kappa z)$
for two forms $\omega$,  $\kappa$ of equal degree in
the differential algebra of the internal spectral triple
$(\aa,\hh,\dd)$. Since the gauge couplings in usual
Yang-Mills theories parameterize gauge invariant
scalar products on the Lie algebra, $z$
deserves the name `non-commutative coupling
constant'. Here is the list of its properties:
\begin{itemize}
\item
$z$ is positive,
\item
$\left[z,\rho(a)\right]=
\left[z,J\rho(a)J^{-1}\right]=0,\qq a\in\aa,$
\item
$[z,\chi]=0,$
\item
$[z,\dd]=0.$
\end{itemize}
For spacetime, $z$ is simply a positive number times
the identity.  For the internal space of the standard
model, the most general $z$ involves $2(1+N)=8$
strictly positive numbers $x,\ y_1,\ y_2,\ y_3,$ $ \tilde
x,\ \tilde y_1,\  \tilde y_2,\ \tilde y_3$,
\bb z&=& \pp{z_w&0\cr 0&\bar z_s},\cr\cr  \cr
&&z_w:=\pp{
x/3\,1_2\ot 1_N\ot 1_3&0&0&0\cr  0&1_2\ot y&0&0\cr
0&0&x/3\,1_2\ot 1_N\ot 1_3&0\cr
0&0&0&y},\cr \cr \cr
&&z_s:=\pp{
\tilde x/3\,1_2\ot 1_N\ot 1_3&0&0&0\cr  0&1_2\ot
\tilde y&0&0\cr
0&0&\tilde x/3\,1_2\ot 1_N\ot 1_3&0\cr
0&0&0&\tilde y},\cr\cr  \cr
&&\qq y:=\pp{
y_1&0&0\cr
0&y_2&0\cr
0&0&y_3},\qq
\tilde y:=\pp{
\tilde y_1&0&0\cr
0&\tilde y_2&0\cr
0&0&\tilde y_3}.\eee
The interpretation of these numbers is
straightforward. The three $y_j$ poise the weak
interactions with the three lepton generations. The
$y_j$ enter independently because the Higgs scalar
couples differently to the three leptons and in
non-commutative geometry the Higgs is part of
the gauge
interactions.  The three $\tilde y_j$ poise the `strong'
interactions with the three lepton generations. They
do not drop out because of the $b$ in $\rho_3$.
However, as we shall see in equations
(\ref{g2}-\ref{g1}), they will only enter as sum: strong
interactions are unbroken and do not generate a Higgs.
$x$ and $\tilde x$ poise weak and strong interactions
with quarks. There is only one number per interaction
because of the Cabbibo-Kobayashi-Maskawa mixing
that we suppose non-degenerate.

In the standard model, the scalars turn out to sit in one
isospin doublet - colour singlet $\varphi$ and their
potential is computed \cite{ks2},
 \bb V(\varphi)=\frac{K}{16L^2}\,|\varphi|^4-
\frac{K}{2L}\,|\varphi|^2.\eee
The coefficients depend on the coupling constants in
$z_w$ only --- because strong interactions do not
contribute to the spontaneous symmetry breaking---
and on squares of the fermion masses, the
Cabbibo-Kobayashi-Maskawa matrix drops out:
\bb K&:=&\frac{3}{2}
\t\left[\left(M_u^*M_u\right)^2\right]x+
\frac{3}{2}
\t\left[\left(M_d^*M_d\right)^2\right]x+
\t\left[M_u^*M_uM_d^*M_d\right]x
+\frac{3}{2}
\t\left[M_e^*M_eM_e^*M_e\,y\right]\cr\cr  &&
-\frac{1}{2} L^2\left[\frac{1}{Nx+\t y}+
\frac{1}{Nx+\t y/2}\right],\label{k}\\ \cr
L&:=&\t\left[M_u^*M_u\right]x+
\t\left[M_d^*M_d\right]x+
\t\left[M_e^*M_e\,y\right]. \label{l}\ee
At the same time, the Yang-Mills Lagrangians for
isospin and colour come out respectively as
\bb
\frac{1}{2}\t\left[\rho\left(F_{2\,\mu\nu},0,0\right)^*
\rho\left({F_2}^{\mu\nu},0,0\right)\,z\right]&=:&
\frac{2}{g^2_2}\,\frac{1}{4}\,\t\left[F_{2\,\mu\nu}^*
{F_2}^{\mu\nu}\right],\qq F_{2\,\mu\nu}
\in \left\{a\in\hhh,\ a^*=-a\right\},
\cr \cr
\frac{1}{2}\t
\left[\rho\left(0,0,F_{3\,\mu\nu}\right)^*
\rho\left(0,0,{F_3}^{\mu\nu}\right)\,z\right] &=:&
\frac{2}{g^2_3}\,\frac{1}{4}\,\t\left[F_{3\,\mu\nu}^*
{F_3}^{\mu\nu}\right],\qq
F_{3\,\mu\nu}\in\left\{c\in M_3(\cc),\
c^*=-c\right\},\eee
with gauge couplings therefore given by
\bb g_2^{-2}&=&{Nx+\t
y}\ ,\label{g2}\\
  g^{-2}_3&=&\frac{4}{3}\,N\tilde
x.\eee
Consequently, we have the following
mass relations:
\bb m_W^2&=&\frac{L}{{Nx+\t y}}\ ,
\label{w}\\ \cr  m_H^2&=&\frac{2K}{L}.\label{h}\ee
So far we have identified the $SU(2)$ of weak isospin
and the $SU(3)$ of colour together with their gauge
couplings. It remains to look at the $U(1)$ of
hypercharge. The Lie algebra of the gauge group is
$\gg=\left\{a\in\aa\ |\ a^*=-a\right\}=
su(2)\op u(1)\op su(3)\op u(1)$. Fortunately, the
hypercharge generator $Y$ is a linear combination of
the two $U(1)$ generators $(0,i,0),\ (0,0,i1_3)$:
\bb Y=\frac{1}{i}\,\rho\left(0,\frac{i}{2},
\frac{i}{6}\,1_3\right).\eee
To compute its gauge coupling,
 we have to recall that $U(1)$ gauge
couplings are conventionally normalized differently
than $SU(n)$ gauge couplings,
\bb \frac{1}{2}\t\left[Y^*Y\,z\right]&=:&
\frac{1}{g^2_1}\,\frac{1}{4}.\eee
Therefore,
\bb g^{-2}_1=Nx+\frac{2}{9}\,N\tilde x+\frac{1}{2}\,
\t y+\frac{3}{2}\t \tilde y.\label{g1}\ee

A final remark of this section concerns the
second, unwanted $U(1)$ which is generated by a
linear combination orthogonal to $Y$. By imposing an
algebraic condition, the Lie algebra $\gg$ is reduced to
the desired subalgebra $su(2)\op u(1)_Y\op su(3)$.
The condition
\bb \t \left[J\rho(1_2,0,0)J^{-1}\,\rho(a,b,c)\right]=0,
\qq (a,b,c)\in\gg, \label{condition}\ee
namely $4N(\t c+\bar b)=0$ selects precisely weak
isospin, hypercharge and colour. This condition looks
like a unimodularity condition because
$J\rho(1_2,0,0)J^{-1}$ is a selfadjoint element in the
commutant of $\rho(\aa)$. Despite this arbitrary
choice of the element in the commutant, the
condition (\ref{condition}) is equivalent to
\bb\t \left[\left(\rho(a,b,c)+J\rho(a,b,c)J^{-1}
\right)P\right]=\t\left[\rho_w(a,b)+\rho_s(b,c)\right]
=0,\eee
 where $P$ is the projection on $\hh_L\op\hh_R$, the
space of particles, and so appears more natural. Note
that this condition is also related to the condition of
vanishing anomalies \cite{anomaly}.

\section{Fuzzy relations among masses and coupling
constants}

\subsection{Masses}

Let us come back to the mass relations
(\ref{w}-\ref{h}). Their coefficients
(\ref{k}-\ref{g2}) contain only squares of
masses and we put
 \bb t:=m_t^2,\qq W:=m_W^2,\qq H:=m_H^2,\qq...\eee
Furthermore, the mass relations are homogeneous in
the variables $x,\ y_1,\ y_2,\ y_3$ and we may set
$x=1/3$ without loss of generality.
Then the two mass relations read
\bb
\frac{H}{W}\,+1&=&\frac{C}{X}\,+3\,\frac{Y}{X}\,
-2\,\frac{X}{1+X},\cr \cr
 X&=&\sum_{j=0}^3\alpha_jy_j \label{xx},\ee
with the following abbreviations
\bb C&:=&\frac{t^2+b^2+c^2+s^2+u^2+d^2}{W^2}\,+
\,\frac{2}{3}\,\frac{tb+cs+ud}{W^2}\,-
\,\frac{1}{3}\,\frac{q^2}{W^2},\cr \cr
q&:=&t+b+c+s+u+d,\cr
\alpha_0&:=&q/3W,\qq\qq
\alpha_1\ :=\ e/W,\qq\qq
\alpha_2\ :=\ \mu/W,\ \qq\qq
\alpha_3\ :=\ \tau/W,\cr
y_0&:=&3x\ =\ 1,\cr\cr
X&:=&\sum_{j=0}^3y_j,\qq\qq
Y\ :=\ \sum_{j=0}^3\alpha_j^2y_j.\eee
An immediate conclusion is that the $W$ mass lies
between the masses of the lightest and the heaviest
fermion, more precisely, if the latter is a quark with
non-degenerate Cabbibo-Kobayashi-Maskawa mixing
in 3 generations, we have
\bb e\,<\,W\,<\,(t+b+c+s+u+d)/3.\eee
 The following lemma justifies the choice of these new
variables $X$ and $Y$, since they
 are independent and bounded.
\begin{description}
\item{\bf Lemma 1.}
{\it
 Let
$\alpha_0,\alpha_1,...,\alpha_N$ be $N+1$ real
numbers, $N\geq 3 $, satisfying the inequalities
$ 0<\alpha_1<...<\alpha_N<1<\alpha_0,$ and
$y_0,\ y_1,\ ...,\ y_N$ be $N+1$ strictly positive
variables. Consider the domain in $\rr^{N+1}$
subject to the constraints
 $y_0=1$ and (\ref{xx}), namely
 \bb D:=\left\{y=(1,y_1,...,y_N),\ y_j>0,\
\sum_{j=0}^N(1-\alpha_j)y_j=0\right\},\eee
and define the variables
$ X:=\sum_{j=0}^Ny_j,\
Y:=\sum_{j=0}^N\alpha_j^2y_j.$
Then,
\begin{enumerate}
\item[i.] $X$ and $Y$ are independent on $D$.
\item[ii.] On $D$, $X$ and $Y$ vary in the open
intervals
\bb
X_{min}:=\frac{\alpha_0-\alpha_1}{1-\alpha_1}&<X<&
\frac{\alpha_0-\alpha_N}{1-\alpha_N}=:X_{max}
,\cr \cr
Y_{min}:=\alpha_0^2+(\alpha_0-1)\,
\frac{\alpha_1^2}{1-\alpha_1}&<Y<&
\alpha_0^2+(\alpha_0-1)\,
\frac{\alpha_N^2}{1-\alpha_N}=:Y_{max}.\eee
\end{enumerate}
}
\end{description}

\noindent{\it Proof.} {\rm i.} follows from a
non-vanishing functional determinant. In fact, it is
sufficient to consider the case $N=3$. We solve the
constraint $\sum_{j=0}^3(1-\alpha_j)y_j=0$:
\bb y_3=-\,\frac{1-\alpha_0}{1-\alpha_3}\,
-\,\frac{1-\alpha_1}{1-\alpha_3}\,y_1
 -\,\frac{1-\alpha_2}{1-\alpha_3}\,y_2,\eee
 we eliminate $y_3$:
\bb X
&=&\,\frac{\alpha_0-\alpha_3}{1-\alpha_3}\,+
\,\frac{\alpha_1-\alpha_3}{1-\alpha_3}\,y_1+
\,\frac{\alpha_2-\alpha_3}{1-\alpha_3}\,y_2,\cr \cr
Y&=&\left(\alpha_0^2-\alpha_3^2\,
\frac{1-\alpha_0}{1-\alpha_3}\right)\,+
\left(\alpha_1^2-\alpha_3^2\,
\frac{1-\alpha_1}{1-\alpha_3}\right)\,y_1+
\left(\alpha_2^2-\alpha_3^2\,
\frac{1-\alpha_2}{1-\alpha_3}\right)\,y_2,
\eee
and compute the functional determinant
\bb\det\,\pp{
\frac{\del X}{\del y_1}&\frac{\del X}{\del y_2}\cr
\frac{\del Y}{\del y_1}&\frac{\del Y}{\del y_2}}\,=\,
\frac{(\alpha_1-\alpha_2)(\alpha_2-\alpha_3)
(\alpha_1-\alpha_3)}{1-\alpha_3}\,\not=0.\eee

\noindent To prove {\rm ii.}, we note that $D$ is
convex and bounded. Indeed, for $j=1,...,N$ we have
\bb (1-\alpha_j)y_j<\,\sum_{j=1}^N(1-\alpha_j)y_j=
-(1-\alpha_{0})y_{0}=\alpha_{0}-1,\eee
 and
$0<y_j<\frac{\alpha_{0}-1}{1-\alpha_j}.$
For every $n=1,...N$, let us define the vector
\bb P_n:=\left(1,0,...,0,
\frac{\alpha_{0}-1}{1-\alpha_n},0,...,0\right)\,
\in\rr^{N+1},\eee
where the $n$ dependent entry
is in the $n$th position. Clearly, the $P_n$ are in the
closure of $D$ and $D$ is the interior of the convex
envelope of the $n$ vectors $P_n$: every $y\in D$
can be written as
\bb y=\sum_{n=1}^N\lambda_n P_n \qq{\rm with}\qq
\lambda_n:=\,\frac{1-\alpha_n}{\alpha_0-1}\
y_n>0\qq {\rm
and}\qq\sum_{n=1}^N\lambda_n=1\eee
because of the constraint. Therefore
\bb X=\sum_{j=0}^Ny_j=\sum_{n=1}^N\lambda_n
\left(1+\frac{\alpha_0-1}{1-\alpha_n}\right)
=\sum_{n=1}^N\lambda_n\,\frac{\alpha_0-\alpha_n}
{1-\alpha_n},\eee
and as $(\alpha_0-\alpha)/(1-\alpha)$ is an
increasing function of $\alpha$,
\bb \frac{\alpha_0-\alpha_1}{1-\alpha_1}\,<X<
\frac{\alpha_0-\alpha_N}{1-\alpha_N}.\eee
Similarly, we obtain the bound on $Y$,
\bb
Y=\sum_{j=0}^N\alpha_j^2y_j=\sum_{n=1}^N\lambda_n
\left(\alpha_0^2+
(\alpha_0-1)\frac{\alpha_n^2}{1-\alpha_n}\right)
\eee
by noting that $\alpha^2/(1-\alpha)$ is increasing
in $\alpha$:
\bb \alpha_0^2+(\alpha_0-1)\,\frac{\alpha_1^2}
{1-\alpha_1}\,<Y<
\alpha_0^2+(\alpha_0-1)\,\frac{\alpha_N^2}
{1-\alpha_N},\eee
ending the proof of the lemma.\vskip 0,5truecm

Note that as $\alpha_0,\ \alpha_1,\ \alpha_N$
vary, $X_{min}$ and $X_{max}$ take all values of
$(1,\infty)$ and $Y_{min}$ and $Y_{max}$ take all
values of $(\alpha_0^2,\infty)$.

Let us now suppose that $C$ is positive. Since
\bb \frac{3}{2}\,W^2C= t^2+b^2+c^2+s^2+u^2+d^2
-(t+b)(c+s+u+d)-(c+s)(u+d),\eee
$C$ is indeed positive in presence of the following
hierarchy of quark masses: $u+d<\min\{c,s\}$,
$c+2s<\min\{t,b\}$.\hfil\break\noindent
 Then the
function of two variables
\bb f(X,Y):=\,\frac{C}{X}\,+3\,\frac{Y}{X}\,
-2\,\frac{X}{1+X}\eee
is decreasing in $X$ and increasing in $Y$ and we get
the following bounds on the Higgs mass:
\bb \tilde H_{min}:= [f(X_{max},Y_{min})-1]W\,<\,H\,<\,
[f(X_{min},Y_{max})-1]W=:H_{max}
,\label{higgsbounds}\ee
but we still must check the positivity of these bounds.

$H_{max}$ is positive. Indeed, $-2X/(1+X)-1>-3$ for
positive $X$ and we shall verify
\bb \left[\alpha_0^2+(\alpha_0-1)\,
\frac{\alpha_N^2}{1-\alpha_N}\right]
\frac{1-\alpha_1}{\alpha_0-\alpha_1}\,>1.\eee
As $\alpha^2/(1-\alpha)$ is increasing in
$\alpha\in [0,1]$, it is sufficient to prove the same
inequality with $\alpha_N$ replaced by $\alpha_1$.
Since
\bb g(\alpha_1):=\,\frac
{\alpha_0^2(1-\alpha_1)+\alpha_1^2(\alpha_0-1)}
{\alpha_0-\alpha_1}\eee
has negative derivative,
$g'(\alpha_1)
=1-\alpha_0<0$,
we obtain $g(\alpha_1)>g(1)=1$.

Concerning the lower bound, we remark that
$\lim_{\tau\nearrow W}\, \tilde H_{min}/W=-3$,
so we have to know when $\tilde H_{min}$ is positive.
\begin{description}
\item{\bf Lemma 2.}
{\it
$\tilde H_{min}$ is positive if and only if
\bb\alpha_N\,<\,\frac{X_+-\alpha_0}{X_+-1}
\label{xplus}\ee
with
$ X_+:=\left(A-1+\sqrt{A^2+10A+1}\right)/6,
\ \ \
A:=C+3\left[\alpha_0^2+(\alpha_0-1)
\alpha_1^2/(1-\alpha_1)\right].$
}
\end{description}
\noindent{\it Proof.}
For $X_{max}\in(1,\infty)$,
$ H_{min}/W=A/X_{max}-
2X_{max}/(1+X_{max})-1$
is positive if and only if $-3X_{max}^2+(A-1)X_{max}
+A>0.$ One root of this polynomial is negative, the
other is $X_+$ and $X_+>1$ because $A>3$. $X_{max}<
X_+$ yields the desired upper bound on $\alpha_N$.
\vskip 0,5truecm

Numerically, for $m_t=176\ GeV$, the bound of
(\ref{xplus}) is $\alpha_3<0.92$, corresponding to
$m_\tau< 76.83\ GeV$.
In particular, since $X_+>\alpha_0^2$ the condition
$ \alpha_3\,<\,\frac{\alpha_0^2-\alpha_0}
{\alpha_0^2-1}\,=\,\frac{\alpha_0}{1+\alpha_0}$
implies positive $\tilde H_{min}$. A fortiori,
$ m_\tau\,<\,\sqrt{Wt/(3W+t)}\,=\, 63\ GeV,\
m_t=\,176\ GeV,$
implies positive $\tilde H_{min}$.

 From now on, we put
\bb H_{min}=\tilde H_{min}\label{tildeh}\ee
 if the latter is positive.

Note that the bounds on the Higgs mass do not depend
on the mass of the intermediate leptons ($m_\mu$).

If $\Delta H:=H_{max}-H_{min}$ denotes the length of
the accessible interval for $m_H^2$, one checks that
\bb \frac{\Delta H}{W}=(\alpha_N-\alpha_1)
(\alpha_0-1) &&\left[\frac{C}
{(\alpha_0-\alpha_N)(\alpha_0-\alpha_1)}\,+3\,
\frac{\alpha_0+\alpha_N-\alpha_0\alpha_N}
{(\alpha_0-\alpha_1)(1-\alpha_N)}\right.\cr \cr
&&\qq\left.+3\,
\frac{\alpha_1+\alpha_N-\alpha_1\alpha_N}
{(\alpha_0-\alpha_N)(1-\alpha_1)}\,+\,
\frac{2}{(\alpha_0+1-2\alpha_N)
(\alpha_0+1-2\alpha_1)}\right].\eee
Therefore the fuzziness disappears if and only if the
sum of the squares of all six quark masses equals
$3m_W^2$ or $m_\tau=m_e$. Indeed, neglecting all
fermion masses but $m_\tau$ and $m_t$,
\bb \Delta m_H:= m_{H\,max}-m_{H\,min}&=&
\left[k \left(\frac{m_\tau}{m_t}\right)^2\,+\,
O\left(\frac{m_\tau}{m_t}\right)^4\,\right]\,m_t,
\cr\cr
k&:=&\frac{\sqrt{3}}{2}\,
\frac{r^3+3r^2-7r-33}{r^3+5r^2+5r-3}\,
\sqrt{r\,\frac{r^2+2r-1}{r+3}},\label{kon}\\ \cr
r&:=&\left(\frac{m_t}{m_W}\right)^2.\label{r}\ee
For $m_t\ =\ 176\ GeV $, we have $k=1.76$.

 If there are only $N=2$ generations of leptons and
quarks (or
likewise three generation of leptons and no quarks)
then the Lemma 1 no longer holds since $Y$ is a
function of $X$. Nevertheless, the Higgs mass varies in
an open interval, an analogue of equation
(\ref{higgsbounds}) holds and $\Delta m_H$ is
governed by the mass difference $m_\mu-m_e$
\cite{iks}. If there is only $N=1$
generation of leptons and quarks (or
two generations of leptons and no quarks) then the
bounds on the Higgs mass collapse, the mass relation
becomes exact, i.e. an equality \cite{ks2}.

\subsection{Coupling constants}

In absence of strong interactions, there is a relation
among the gauge couplings $g_1$ and $g_2$ \cite{is2}
because then, only $x$ and $\t y$ appear. Depending
on the fermion content, this relation is exact or fuzzy.
If there are only quarks in any number of
generations, we
 have $\sin^2\theta_w=1/5$, and for only
leptons in any number of generations,
$\sin^2\theta_w=1/3$. For leptons {\it and} quarks the
relation becomes fuzzy,
\bb 1/5<\sin^2\theta_w=\,\frac{x+\t
y}{5x+3\t y}<1/3.\eee
However, without strong interactions, the geometric
version of the standard model leads to wrong electric
charges, up and down quark with opposite charges or
charged neutrinos.

The addition of strong interactions cures this problem
and introduces two more parameters, $\tilde x$ and
$\t \tilde y$. Consequently
\bb \sin^2\theta_w=\frac{g_2^{-2}}{g_1^{-2}+g_2^{-2}}
=\frac{Nx+\t y}{2Nx+\frac{2}{9}\,N\tilde x
+\frac{3}{2}\,\t y+\frac{3}{2}\,\t\tilde y}\eee
is only bounded from above,
\bb\sin^2\theta_w<\frac{2}{3}\,\left(1+\frac{1}{9}
\left(\frac{g_2}{g_3}\right)^2\right)^{-1}.\eee
Note that the addition of right-handed neutrinos to
the standard model \cite{gb} improves this
constraint,
\bb\sin^2\theta_w<\frac{1}{2}\,\left(1+\frac{1}{12}
\left(\frac{g_2}{g_3}\right)^2\right)^{-1}.\eee
Using
\bb\alpha_{em}:=\,\frac{g_{em}^2}{4\pi}\,=\,
\frac{g_2^2\sin^2\theta_w}{4\pi}\eee
 we rewrite the inequality (\ref{sinth}):
\bb \alpha_3:=\,\frac{g_3^2}{4\pi}\,>\,
\frac{\alpha_{em}}{6(1-3/2\,\sin^2\theta_w)}.\eee
It now says that the strong fine structure constant
cannot be very small, $\alpha_3\,>\,0.002$.
Experimentally it is around 0.11.

\section{Conclusions}

Non-commutative geometry explains
the constraint
$ m_Z=m_W/\cos\theta_w.$
Although being an equality it is stable under
renormalization flow. Non-commutative geometry
has three additional constraints: (\ref{wth})
explains why the top is so heavy, (\ref{hth})
predicts the Higgs mass  and (\ref{sinth})
constrains the weak mixing angle and the strong
coupling constant. All three constraints are fuzzy, i.e.
given by inequalities and therefore locally stable
under renormalization flow \cite{agm2}.
Local stability should be sufficient since the theory
contains no superheavy particle.
 Numerically
the Higgs mass is predicted at $m_H\ =\ 280\pm33\
GeV$ for the current top mass of $m_t\ =\ 176\pm18\
GeV$, a prediction to be tested within ten years.
\vskip 1truecm

\noindent
It is as pleasure to acknowledge helpful
advice of Alain Connes and Gilles Esposito-Far\`ese.

\vfil\eject

\vskip 2truecm

{\bf Figure caption:} Lower and upper bounds of the
Higgs mass as a function of the top and $\tau$ masses,
all other masses being set to their experimental values.
For the experimental value, $m_\tau\ =\ 1.8\ GeV$, the
two bounds differ by around $10^{-2}\ GeV$ in the
indicated range of $m_t$.

 
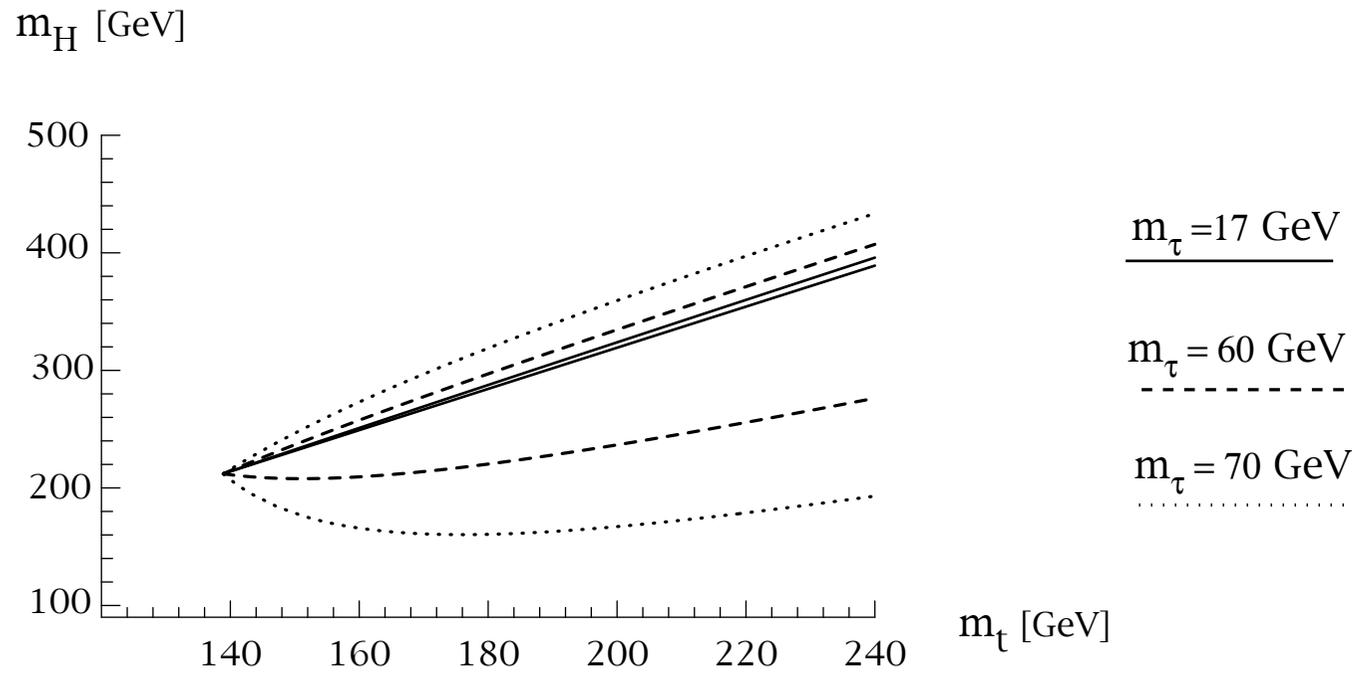
\end{document}